\documentclass[aps,10pt,pre,amsmath,amsfonts,superscriptaddress,nofootinbib,twocolumn]{revtex4-1}
\usepackage{mathrsfs}
\usepackage{graphicx,color,float}
\usepackage{bm,bbm}
\usepackage{physics}
\usepackage{multirow}
\usepackage{hyperref}
\usepackage[caption=false]{subfig}
\usepackage{algorithm,algpseudocode}
\def\?#1{}
\def\OO{\mathcal{O}}
\DeclareMathOperator{\erfc}{erfc}

\begin{document}

\title{Dynamics and fluctuations of minimally-structured glass formers}

\author{Patrick Charbonneau}
\affiliation{Department of Chemistry, Duke University, Durham, North Carolina 27708, USA}
\affiliation{Department of Physics, Duke University, Durham, North Carolina 27708, USA}
\author{Yi Hu}
\thanks{Corresponding author: \?}
\email{yi.hu@duke.edu}
\affiliation{Department of Chemistry, Duke University, Durham, North Carolina 27708, USA}
\author{Peter K.~Morse}
\affiliation{Department of Chemistry, Duke University, Durham, North Carolina 27708, USA}
\affiliation{Department of Chemistry, Princeton University, Princeton, New Jersey 08544, USA}
\affiliation{Department of Physics, Princeton University, Princeton, New Jersey 08544, USA}
\affiliation{Princeton Institute of Materials, Princeton University, Princeton, New Jersey 08544, USA}

\date{\today}

\begin{abstract}
The mean-field theory (MFT) of simple glasses, which is exact in the limit of infinite spatial dimensions, $d\rightarrow\infty$, offers theoretical insight as well as quantitative predictions about certain features of $d=3$ systems. In order to more systematically relate the behavior of physical systems to MFT, however, various finite-$d$ corrections need to be accounted for. Although some efforts along this direction have already been undertaken, theoretical and technical challenges hinder progress. A general approach to sidestep many of these difficulties consists of simulating minimally-structured models whose behavior smoothly converges to that described by the MFT as $d$ increases, so as to permit a dimensional extrapolation. Using this approach, we here extract the small fluctuations around the dynamical MFT captured by a standard liquid-state observable the non-Gaussian parameter $\alpha_2$. The results provide insight into the physical origin of these fluctuations as well as a quantitative reference with which to compare observations in more realistic glass formers.
\end{abstract}

\maketitle

\section{Introduction}
Over the last fifteen years, the mean-field theory (MFT) of simple glasses has been steadily worked out~\cite{parisi2010mean,charbonneau2017glass,francesco2020theory}. In some respects, the approach, which is exact when the number of spatial dimensions diverges, $d\rightarrow\infty$, has met stunning success. It captures certain aspects of finite-$d$ jamming physics with remarkable accuracy and predicts a novel type of (Gardner) transition in amorphous solids that has found various experimental echoes~\cite{seguin2016experimental,hammond2020experimental,xiao2022probing,kool2022gardner}. From a theoretical physics standpoint, it has also brought descriptions that were based on loose physical analogies or uncontrolled approximations under a single coherent umbrella. For instance, it has confirmed the validity of the random first-order transition (and associated scenarios) in the limit $d\rightarrow\infty$ as well as the mean-field nature of the mode-coupling theory of glasses, despite its quantitative failings~\cite{ikeda2010mode,pihlajamaa2023unveiling}.

The MFT of simple glasses, however, is no panacea. By construction, it neglects fluctuations and only accounts for minimal pair-wise correlations. (Including higher-body correlations, which vanish as $d\rightarrow\infty$, would require an even richer--and more unwieldy--cluster-based approach~\cite{liu2021dynamics}.) These simplifications obfuscate the possible connection between MFT and the structurally and dynamically rich behavior of $d=3$ glass formers. 
Early efforts to estimate the impact of finite-dimensional pair structure correlations on the MFT-like caging transition have met only limited success~\cite{mangeat2016quantitative,charbonneau2022dimensional}. More recently, the extent and nature of small caging fluctuations have also been more successfully considered~\cite{biroli2022local,folena2022equilibrium}. Large (instantonic) deviations, however, remain out of reach even for single-particle processes~\cite{biroli2020unifying}. 

The dynamical MFT (DMFT)~\cite{maimbourg2016solution,szamel2017simple} also remains challenging to hammer out. Its analytical formulation is so complex that only under a limited range of conditions can it be evaluated~\cite{manacorda2020numerical,manacorda2022gradient}. Out-of-equilibrium conditions are particularly thorny~\cite{agoritsas2018out,agoritsas2019out1,agoritsas2019out2,morse2021direct}. To tease out some of the out-of-equilibrium physics, recent efforts have taken to simulating minimally-structured models of glasses in finite $d$ and extrapolating the results to the limit $d\rightarrow\infty$~\cite{manacorda2022gradient,charbonneau2023jamming}. In these studies, removing structural complexities ensured that (non-instantonic) corrections scale perturbatively in $1/d$ and can therefore be reasonably controlled through dimensional extrapolation. 

Although computing small dynamical fluctuations around the DMFT also remains out of theoretical reach, a first-principle understanding of the origin of single-particle contributions has recently been worked out~\cite{biroli2022local}. The observable that naturally describes small fluctuations in the caging regime, the (single-particle) dynamical susceptibility $\chi_4(t)$, can indeed be immediately related with an observable commonly used to characterize supercooled liquids~\cite{kob1997dynamical,yamamoto1998heterogeneous,berthier2011theoretical}, the non-Gaussian parameter $\alpha_2(t)$, both of which diverge around the caging transition in $d\rightarrow\infty$. From a different standpoint, the DMFT describes the single-particle equilibrium dynamics as a Langevin equation with a purely Gaussian (albeit colored) noise, which can be integrated to obtain the mean squared displacement (MSD), $\Delta(t)$~\cite{maimbourg2016solution,szamel2017simple}. Deviations of the self-van Hove function from Gaussianity are therefore necessarily finite-$d$ corrections. In order to determine if $d=3$ observations of $\alpha_2(t)$ in simple glass formers in any way reflect fluctuations around the DMFT, however, these fluctuations should first be quantified.

Here, building on the approach developed to study the out-of-equilibrium DMFT, we extract the $d\rightarrow\infty$ behavior of $\alpha_2(t)$ by studying and comparing two minimally-structured models, the random Lorentz gas (RLG) and the Mari-Kurchan (MK) model, that separately converge to the MFT of simple glasses. \emph{En passant}, we validate the DMFT description of the MSD as well as the $d\rightarrow\infty$ small caging fluctuations.

\section{Models and Mapping}
\label{sec:models}
The RLG is a point tracer that evolves within the void (empty) space left by Poisson distributed spherical obstacles of unit diameter and scaled density $\hat\varphi = \rho V_d/d$, where $\rho$ and $V_d$ denote the point intensity (number density) and the $d$-dimensional unit radius sphere volume, respectively. It is here simulated as in Refs.~\cite{biroli2020unifying,charbonneau2020percolation,biroli2022local}. Initial configurations are obtained using a cavity reconstruction scheme~\cite{biroli2020unifying,biroli2022local}. In the caging regime, for densities above the MFT dynamical transition at which glass-like (and mode coupling theory-like) caging emerges, $\hat\varphi\gtrsim \hat\varphi_\mathrm{d}$, the tracer is placed at the origin and surrounded by a spherical shell of obstacles that precludes overlaps with the tracer and is of sufficient thickness to prevent its escape over the simulation time. In the diffusive regime, for $\hat\varphi \lesssim \hat\varphi_\mathrm{d}$, initial configurations are obtained through quiet planting~\cite{krzakala2009hiding,biroli2020unifying}. This approach is implemented by placing the tracer at the center of a simulation box, and then distributing obstacle positions uniformly at random within that box, rejecting any obstacles that cover the origin. A typical simulation box consists of $10^5$ obstacles for $d=3$, and up to $10^6$ obstacles for $d=24$ (under Leech lattice symmetry~\cite{conway1986soft,cohn2017sphere,vardy1993maximum,van2016cryptographic}, see Appendix~\ref{appd:box}). Results are averaged over at least $10^3$ independent realizations of obstacle positions.

The Mari-Kurchan (MK) model~\cite{mari2009jamming,mari2011dynamical} consists of $N$ hard spherical particles of unit diameter interacting through shifted pair interactions. It is simulated as in Ref.~\cite{charbonneau2014hopping}. Here again, initial configurations are obtained by quiet planting~\cite{charbonneau2014hopping}. After placing particles uniformly at random within the simulation box, pairwise shifts are sampled uniformly at random within that box, but values that result in the two particles overlapping are rejected. Like for the RLG, a scaled density can be defined $\hat\varphi = 2^d\rho V_d/d$, where the factor of $2^d$ accounts for the different obstacle diameter conventions used in the two models.

As described in Refs.~\cite{biroli2020unifying,biroli2020mean,manacorda2022gradient}, in the $d\rightarrow\infty$ limit, the two models only differ by a trivial scaling factor,
\begin{equation} \label{eq:mapping}
\begin{aligned}
2 \hat\varphi_\mathrm{RLG} &\leftrightarrow \hat\varphi_\mathrm{MK}\\
\hat\Delta_\mathrm{RLG} &\leftrightarrow 2 \hat\Delta_\mathrm{MK}\\
[ \langle \hat{r}^4(t) \rangle ]_\mathrm{RLG} & \leftrightarrow 4 [ \langle \hat{r}^4(t) \rangle ]_\mathrm{MK},\\
\end{aligned}
\end{equation}
where $\hat\Delta(t)=d\Delta(t)$ for the MSD $\Delta(t)=[ \langle r^2(t) \rangle ]$ and $[ \langle \hat{r}^4(t) \rangle ]=d^2[ \langle r^4(t) \rangle ]$ for the mean quartic displacement, after thermal $\langle ... \rangle$ averaging (over displacements within a cage) and disorder $[\ldots]$ averaging (over cage realizations). (For notational convenience, in the rest of the text we omit the subscript (RLG or MK) when the discussion applies to both models or when the model is clearly specified in the surrounding text.) The scaling of time depends both on the model and on the chosen dynamics (either Newtonian~\cite{skoge2006packing,charbonneau2014hopping} or Brownian~\cite{foffi2005scaling,scala2007event}, see Appendix~\ref{appd:microdynamics}). The mean tracer velocity, which controls time scaling of the RLG, can indeed be chosen arbitrarily. For convenience, the scaled time for the RLG is here aligned with that of the MK model (and with the reference DMFT calculation),
\begin{equation}
\hat t_\mathrm{RLG} \leftrightarrow \hat t_\mathrm{MK}.
\end{equation}
When the tracer mean velocity is set to unity, this choice leads to $\hat t = \sqrt{d} t, d t / 2$ for the Newtonian and Brownian RLG, and $\sqrt{2} d t$ for the Newtonian MK model.

By contrast, no universal mapping is possible for the the dynamical susceptibility $\chi_4(t)=[ \langle r^4(t) \rangle ]-\Delta^2(t)$ (and the scaled $\hat\chi_4= d( [ \langle \hat{r}^4(t) \rangle ] - \hat\Delta^2) = d^3 \chi_4$~\cite{biroli2022local}). The two models are nevertheless expected to correspond in certain limits (see Appendix~\ref{appd:mapping}),
\begin{equation} \label{eq:mapping2}
\hat\chi_\mathrm{4,RLG} \leftrightarrow
\begin{cases}
4 \hat\chi_\mathrm{4,MK}, & \text{short times \& long-time diffusion,} \\
8 \hat\chi_\mathrm{4,MK}, & \text{long-time caging.}
\end{cases}
\end{equation}

\begin{figure*}
\centering
\includegraphics[width=1\textwidth]{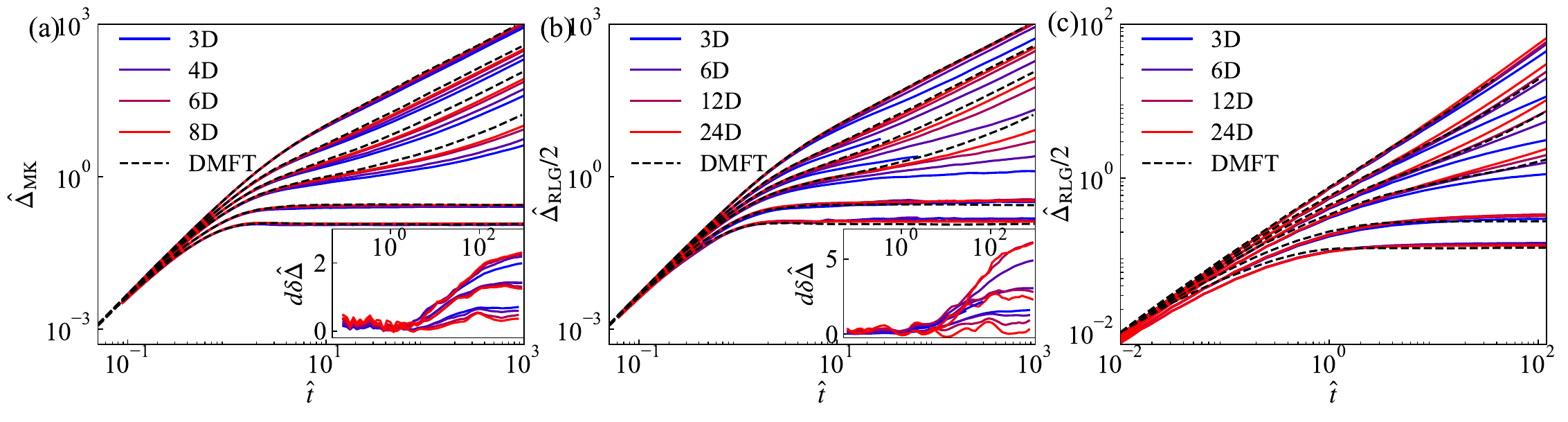}
\caption{MSD for (a) the MK model at $\hat\varphi=1,2,3,4,6,8$ (from top to bottom) with Newtonian dynamics and for the RLG at corresponding $\hat\varphi=0.5,1,1.5,2,3,4$ (from top to bottom) with (b) Newtonian and (c) Brownian dynamics.
Numerical solution of the DMFT equations are also included (dashed black lines)~\cite{manacorda2020numerical}. (insets)
Rescaled deviation of simulation results from the DMFT prediction for Newtonian dynamics with (a) $\hat\varphi=1,2,3$ and (b) $\hat\varphi=0.5$, 1 and 1.5 (from bottom to top). In all cases, a scaling collapse is achieved as $d$ increases, but the correction is smaller for the MK model than for the RLG at corresponding densities. DMFT predictions for Brownian dynamics deviate too strongly from finite-$d$ results for a similar collapse to be attempted.}
\label{fig:msdnewbr}
\end{figure*}

\section{MSD time evolution}
\label{sec:MSDtime}

In order to validate the perturbative nature of these models, the MSD from numerical simulations is compared with the DMFT prediction of Ref.~\onlinecite{manacorda2020numerical} for $d\rightarrow\infty$ hard spheres. Results for the MK model are expected to naturally converge to that limit as $d$ increases, and those for the RLG to do so after Eq.~\eqref{eq:mapping} rescaling. For the RLG, an additional effect must be taken into account. As carefully discussed in Ref.~\cite{biroli2020unifying}, for $d\lesssim8$ the percolation transition takes place at densities for which MFT predicts that the tracer should diffuse~\cite{biroli2020unifying}. Because percolation physics is not perturbative relative to the MFT of simple glasses, results in its vicinity do not smoothly converge to the DMFT prediction. Therefore, for $\hat\varphi_\mathrm{p}<\hat\varphi<\hat\varphi_\mathrm{d}$, we only consider the short-time MSD, i.e., for $\hat{t}\ll D^{-1}$, where $D$ is the diffusivity constant. 

Once these effects are accounted for, the dimensional evolution of the MSD is smooth for both the RLG and the MK model (Fig.~\ref{fig:msdnewbr}). For systems with Newtonian dynamics, the trend is robust down to $d=3$ . Low and high density results exhibit only small deviations from the DMFT predictions. As expected from studies of caging susceptibility~\cite{biroli2022local,folena2022equilibrium}, significantly larger deviations emerge upon approaching $\hat\varphi_\mathrm{d}$. In all cases, the deviations are consistent with perturbative corrections (Fig.~\ref{fig:msdnewbr} insets),
\begin{equation} \label{eq:reldev}
\delta \hat\Delta(\hat{t}) = \frac{|\hat{\Delta}_\mathrm{MFT}(\hat{t}) - \hat{\Delta}(d,\hat{t})|}{\hat{\Delta}_\mathrm{MFT}(\hat{t})} \sim 1/d.
\end{equation}
For $\hat\varphi\ll\hat\varphi_\mathrm{d}$, the collapse is nearly quantitative, but higher-order corrections can be detected for $d=3$ at the highest (diffusive) density considered. In all cases, the overall correction is about twice as large for the RLG relative to the MK model.

By contrast, Brownian dynamics simulations deviate from DMFT predictions at all times. They are overestimated at short times $\hat t\lesssim 1$ and underestimated at long times $\hat t \gtrsim 10^1$. For the former, the analytical expansion of the MSD reveals that the associated discrepancy follows from numerical imprecision in solving the DMFT equations~\cite{manacorda2020numerical} (see Appendix~\ref{appd:brshorttime}). For the latter, Ref.~\cite{manacorda2020numerical} noted that although the  equations are exact in the $d \rightarrow \infty$ limit, a singularity in the memory function gives rise to severe numerical integration challenges. In the caging regime, this difficulty was found to result in the long-time limit of the MSD from the DMFT differing from the (independent) static MFT analysis~\cite{manacorda2020numerical}. Although no similar comparison can be made in the diffusive regime, one can reasonably expect the same underlying issue to be at play there as well. 

\section{MSD long-time scaling}
\label{sec:MSDcage}

\begin{figure*}
\includegraphics[width=1\textwidth]{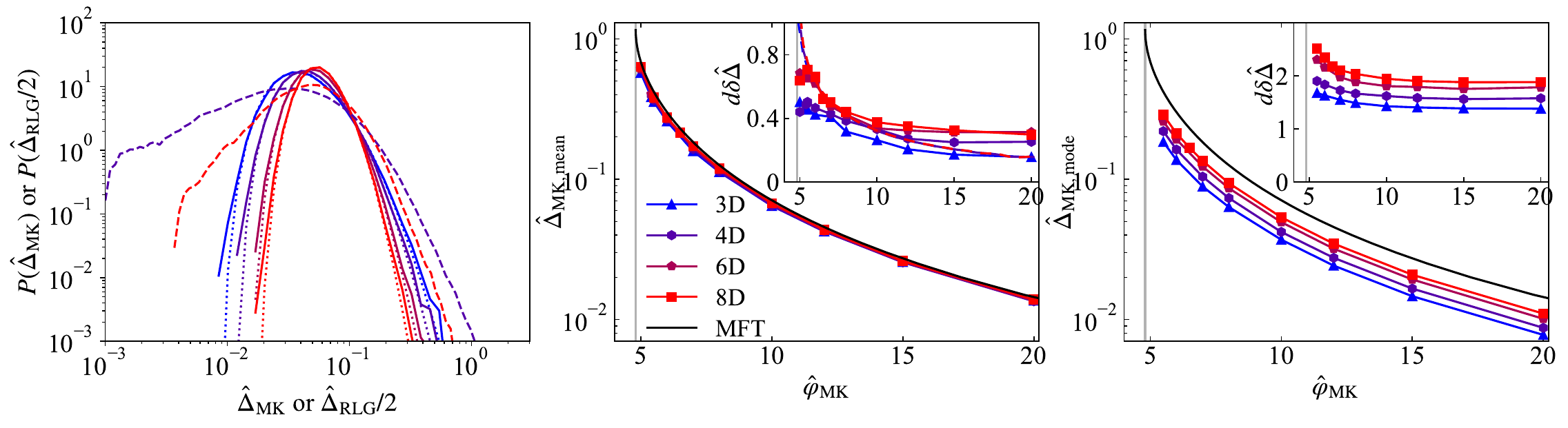}
\caption{(a) Distribution of cage sizes for the MK model in $d=3,4,6,8$ (from blue to red solid line) fit with a log-normal distribution (dotted lines), and compared with the wider distribution obtained in the RLG in $d=4,8$ (dashed lines). (b) Mean and (c) modal cage size for the MK model in finite $d$ (points) compared with the $d\rightarrow\infty$ MFT predictions (black line) above $\hat\varphi_\mathrm{d}=4.81...$ (gray line). Simulation results only slightly undershoot the theoretical curve at all densities. 
(inset) Perturbative $1/d$ correction of the simulation results. In (b, inset), Gaussian ansatz estimate for these corrections is included as reference (dashed line), which ends at a finite value beyond the plot range (e.g. 2.5 for $d=8$).}
\label{fig:mkplateau}
\end{figure*}

As further validation of the perturbative nature of these models, we consider the long-time MSD in the caging regime (Figs.~\ref{fig:mkplateau} and \cite[Fig.~3]{biroli2020mean}). Here again, a smooth dimensional evolution is observed. An interesting difference between the finite-$d$ RLG and MK model is nevertheless noted.
While both systems have $\hat \Delta(d, \hat{t}) < \hat \Delta_\mathrm{MFT}(\hat t)$ in the diffusive regime, deviations are of opposite sign in the caging regime. 
In Ref.~\cite{biroli2020mean}, the sign of the deviation was found to differ between the mean and the modal squared displacements of the RLG, as a result of a heavy tail in the probability distribution of cages in that system. The absence of such a distinction in the MK model suggests a significantly smaller cage size anisotropy. To quantify this effect, we repeat for the MK model the RLG analysis in Ref.~\onlinecite{biroli2020mean} for the long-time plateau [$\hat\Delta=\hat\Delta(\hat t\rightarrow\infty)$] of both mean and modal cage size (Fig.~\ref{fig:mkplateau}). We also follow Ref.~\cite[Fig.~2(b)]{biroli2022local} in fitting the cage sizes distribution with a log-normal form, which describes the distribution well even for $d=3$. By contrast, the large tail observed in the RLG model at corresponding $d$ and $\hat{\varphi}$ markedly deviates from a log-normal form. In other words, cages for the MK model are more narrowly distributed than for the RLG (see Eq.~\eqref{eq:mapping}) and seemingly less affected than the RLG by non-perturbative (instantonic) hopping effects~\cite{charbonneau2014hopping,biroli2020unifying}.

As expected from an earlier report~\cite{charbonneau2014hopping}, the cage size--as determined from both the modal and the mean squared displacements--closely follows the MFT prediction for all $d$ considered nearly down to $\hat\varphi_\mathrm{d}$. Its predicted square-root singularity at $\hat\varphi_\mathrm{d}$, however, is expected to be rounded by hopping in finite $d$~\cite{charbonneau2014hopping}. Here again, these processes are less prevalent in the MK model than in the RLG. For instance, for $\hat\varphi=5.5$, $\hat\Delta(\hat t = 10^4)/\hat\Delta(\hat t = 10^2)=1.002$ which results in an difference in $d \delta \hat \Delta$ of only about $1\%$, while for the RLG a comparable convergence is only possible for $2\hat\varphi_{\mathrm{RLG}} \ge 6$~\cite{biroli2020mean}.  A ``long-time'' plateau can therefore be identified in the MK model at lower $\hat\varphi$ without resorting to the modal displacement to screen away the fat tail of large displacements. From this standpoint, cages in the MK model are clearly better defined than in the RLG.

A consideration of the perturbative regime more specifically finds that although higher-order corrections are noticeable in $d=3$ and $4$, the $1/d$ collapse appears converged by $d=6$-$8$ at high $\hat\varphi$ (Fig.~\ref{fig:mkplateau} insets), which is also consistent with fluctuations being relatively small in this model. Although the perturbative prefactor grows upon approaching $\hat\varphi_\mathrm{d}$ in a way consistent with a divergence, the available $d$ range is insufficient to probe that phenomenon directly. Results for $\hat{\varphi}=5$ indeed already deviate from the expected scaling in the higher $d$ attainable. As in Refs.~\cite[Fig.~3]{biroli2020mean} and \cite[Fig.~2d]{charbonneau2014hopping}, we can nevertheless compare the $1/d$ corrections with those from the Gaussian ansatz. While for the RLG this ansatz was far off the mark~\cite{biroli2020mean}, for the MK model the results are more mixed. The prediction significantly underestimates deviations at higher $\hat\varphi$, as it does for the RLG, but it describes reasonably well the regime $5.5 \lesssim \hat\varphi \lesssim 10$, where a scaling collapse is observed. Because the Gaussian ansatz does not lead to a divergence at $\hat\varphi_\mathrm{d}$, however, deviations are expected to grow more pronounced as $\hat{\varphi}$ is further decreased toward the dynamical transition. The agreement of the MSD cage in Fig.~\ref{fig:mkplateau}b (inset) and \cite[Fig.~2d]{charbonneau2014hopping} should therefore be understood as largely fortuitous. In any event, this overall analysis confirms the perturbative nature of the two minimally-structured models considered here.

\section{Dynamical fluctuations in the diffusive regime}
\label{sec:alpha2}

\begin{figure}
\centering
\includegraphics[width=1\columnwidth]{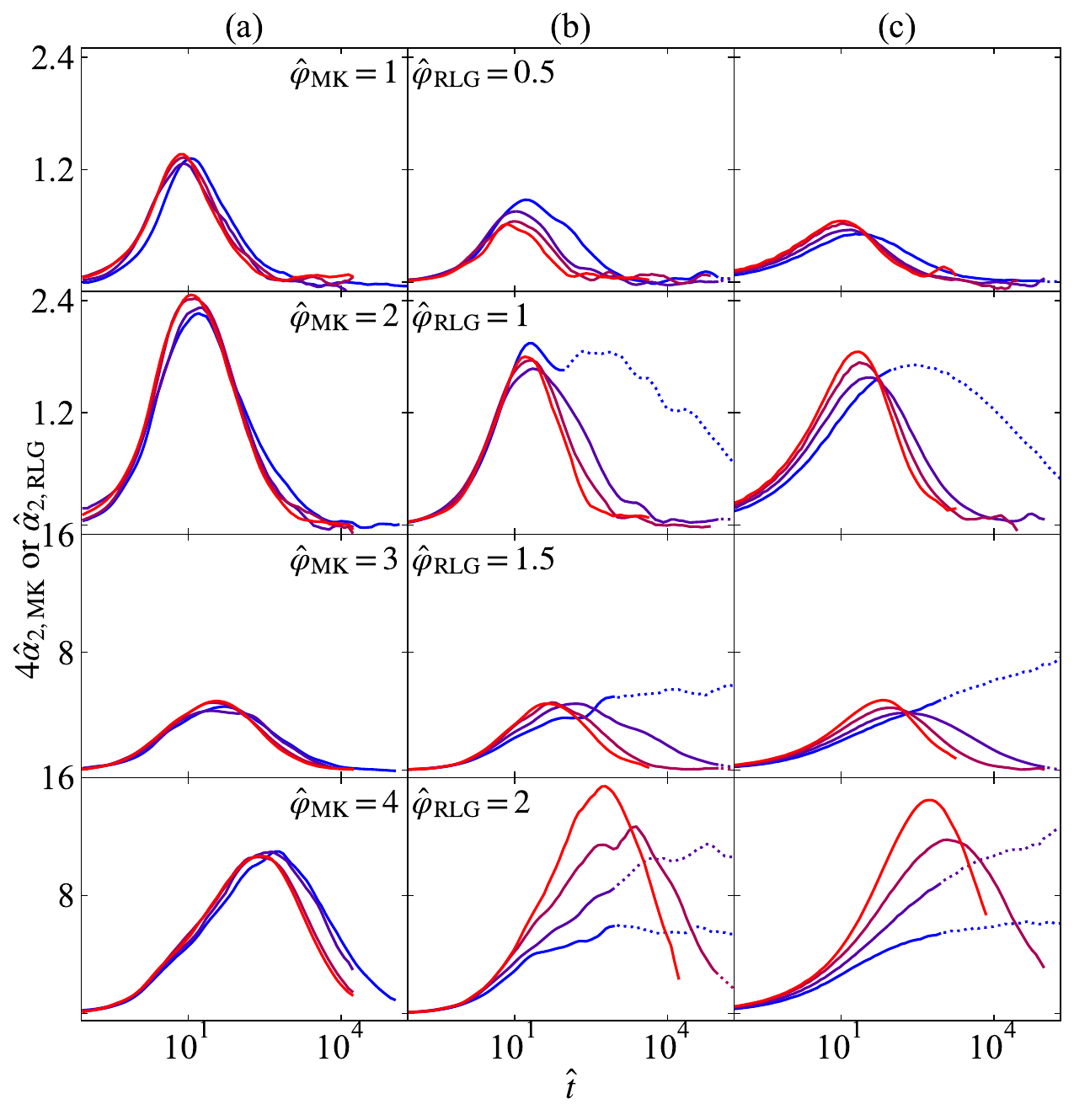}
\caption{Evolution of $\hat\alpha_2$ for (a) the MK model for $d=3,4,6,8$ (blue to red lines) at $\hat\varphi=1,2,3,4$ and the RLG with (b) Newtonian and (c) Brownian dynamics for $d=3,6,12,24$ (blue to red lines) at corresponding $\hat\varphi=0.5, 1, 1.5, 2$. Both systems exhibit a single peak that grows and moves to long times as $\hat\varphi \rightarrow \hat\varphi_\mathrm{d}^-$. (Note that the ordinate axis uses a different scale in the upper and lower panels.) While $\hat\alpha_2$ presents a similar peak compared to the MK model, percolation physics intervenes in the RLG at long times for the larger densities. The resulting deviations are specially marked at small $d$ (dotted out curves), resulting in $\hat{\alpha}_2$ peaking at much longer times.}
\label{fig:mkalpha2time}
\end{figure}

In order to characterize finite-$d$ corrections to the DMFT more systematically, we consider the dynamical fluctuations around the theoretical predictions. Different observables for capturing single-particle fluctuations are commonly used, depending on whether the caging or diffusive regime is considered. In the former, one typically considers the scaled four-point susceptibility, i.e., the kurtosis,
\begin{equation} \label{eq:chi4def}
\left( \frac{\chi_4}{\Delta^2} \right) (t) = \frac{[ \langle r^4(t) \rangle ]}{[ \langle r^2(t) \rangle ]^2} - 1.
\end{equation}
which plateaus at long times. In the latter, because $\chi_4(t)$ diverges at long times we instead consider the non-Gaussian parameter~\cite{biroli2022local}, 
\begin{equation}
\label{eq:alpha2def}
\alpha_2(t) = \frac{d}{d+2} \frac{[ \langle r^4(t) \rangle ]}{[ \langle r^2(t) \rangle ]^2} - 1.
\end{equation}
which vanishes at short and long times and peaks in between, and is dimensionally rescaled as $\hat\alpha_2(\hat t) = d \alpha_2(\hat t)$~\cite{fnote:a2scaling}. Note that the two observables are also linearly related at constant $d$, 
\begin{equation} \label{eq:a2chimap}
\hat\alpha_2 = \frac{d}{d+2} (\frac{\hat\chi_4}{\hat\Delta^2}  - 2),
\end{equation}
which in the limit $d \rightarrow \infty$ simplifies to  $\hat\alpha_2 = \hat\chi_4/\hat\Delta^2 - 2$. Note also that the full time evolution of $\hat\chi_4(\hat t)$, like that of $\hat\chi_4(\hat t \rightarrow \infty)$,  does not simply map between the MK model and RLG, as further discussed in Sec.~\ref{sec:chi4}.

The time evolution of $\hat{\alpha}_2$ at various $\hat\varphi < \hat\varphi_\mathrm{d}$ is shown in Fig.~\ref{fig:mkalpha2time}. The results are qualitatively reminiscent of what has long been reported in supercooled liquids~\cite{kob1997dynamical,yamamoto1998heterogeneous,berthier2011theoretical}, and, as expected~\cite{gleim1998does}, both Newtonian and Brownian dynamics give rise to fairly similar curves as density increases. In general, the growth of $\alpha_2$ as $\hat\varphi$ increases is consistent with cage escapes then proceeding through low-dimensional pathways within which motion along many directions is severely constrained, such as transport through tunnel-like openings~\cite{biroli2022local}.

Before considering the simulation results in more detail, however, an important caveat must be made. Because $\alpha_2$ diverges at the \emph{non-perturbative} percolation transition of the RLG~\cite{hofling2008critical,biroli2020unifying}, results for $\hat\varphi\gtrsim \varphi_\mathrm{p}$ must be neglected in order to assess the \emph{perturbative} physics associated with small fluctuations around MFT. 

Once that is done, $\hat\alpha_2(\hat t)$ presents a single peak at nearly the same time for all $d$, and at low densities the full curves collapse reasonably well. Upon approaching $\hat\varphi \rightarrow \hat\varphi_\mathrm{d}$, however, peaks no longer collapse and instead grow with $d$. What gives rise to these deviations? Recall that MFT (for $d\rightarrow\infty$) finds that this peak should diverge for $\hat\varphi \rightarrow \hat\varphi_\mathrm{d}^{+}$, and that a similar singularity is expected for $\hat\varphi \rightarrow \hat\varphi_\mathrm{d}^{-}$~\cite{biroli2022local}. In finite $d$, two phenomena could be at play: (i) activated events turn any MFT divergences into a crossover in finite $d$~\cite{charbonneau2017glass}; and (ii) the number of directions along which particle motion can be correlated is finite, thus bounding the growth of $\alpha_2$. 

\begin{figure}
\centering
\includegraphics[width=1\columnwidth]{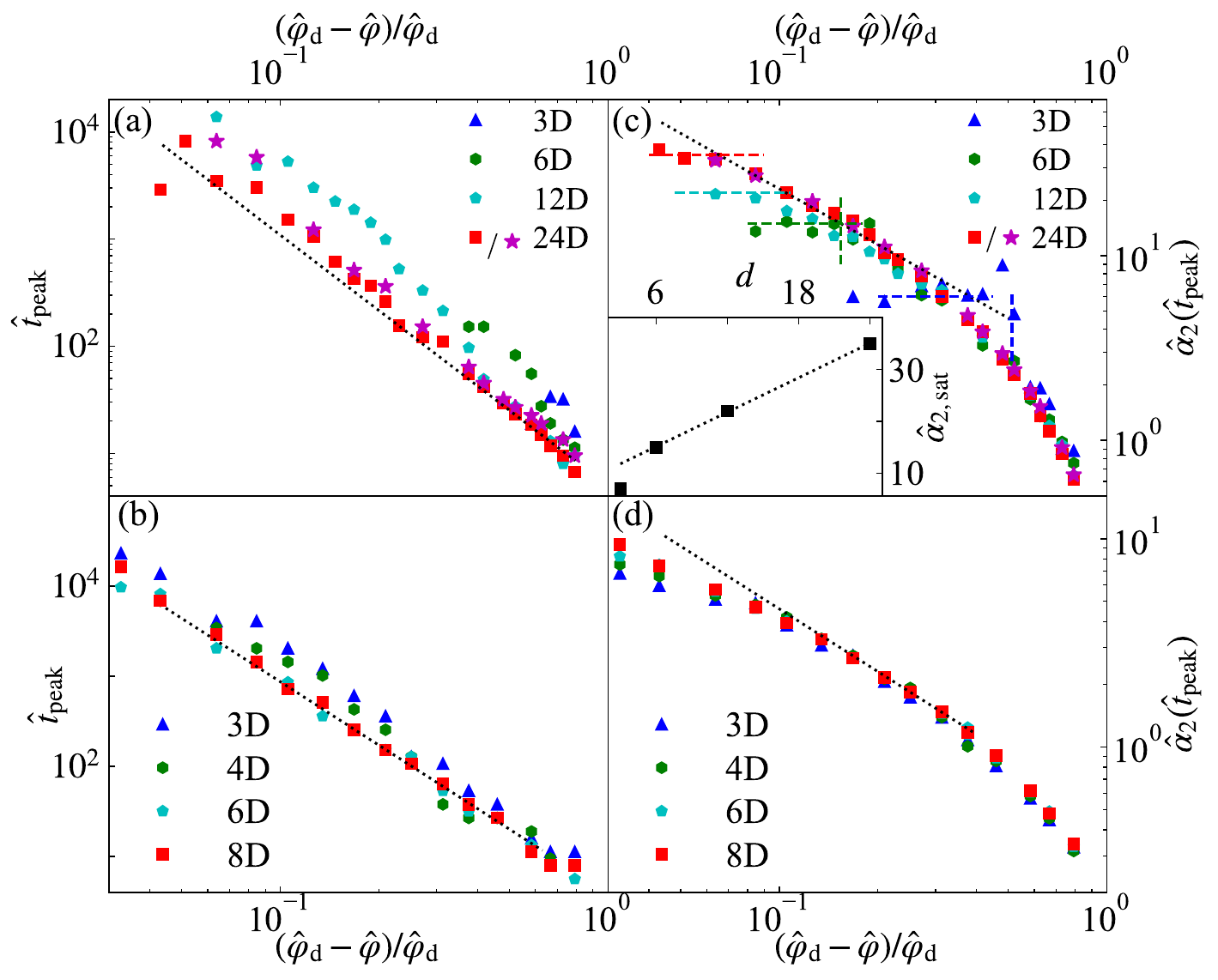}
\caption{Peak characteristics for $\hat\alpha_2(\hat t)$. Growth of the peak time for  (a) the RLG and (b) the MK model as $\hat\varphi$  approaches $\hat\varphi_\mathrm{d}$ for different $d$. In both systems the results are roughly consistent with the expected critical scaling $\hat t_\mathrm{peak}  \sim (\hat\varphi_\mathrm{d} - \hat\varphi)^{-\gamma}$ with $\gamma=2.34\ldots$ (black dotted line). Evolution of the peak height for (c) the RLG and (d) the MK model as $\hat\varphi$ approaches $\hat\varphi_\mathrm{d}$. The growth first follows the expected critical scaling $\hat \alpha_2(\hat t_\mathrm{peak}) \sim (\hat\varphi_\mathrm{d} - \hat\varphi)^{-1}$ (black dotted line), but eventually saturates. In (c) results deviate from this trend around the percolation threshold ($\hat\varphi_\mathrm{p}$ for $d=3, 6$ are marked as vertical dashed lines). (inset) The saturation plateau of $\hat\alpha_2(\hat t_\mathrm{peak})$ grows roughly linearly with $d$ (dotted line).}
\label{fig:a2peakscaling}
\end{figure}

A quantification of peak characteristics offers some insight into which phenomenon dominates in this regime. First, consider the peak time. Peak non-Gaussianity coincides with cage escape and is therefore expected to follow the growth of the structural relaxation time. We therefore expect a scaling $\hat t_\mathrm{peak} \sim (\hat\varphi_\mathrm{d} - \hat\varphi)^{-\gamma}$ with a non-universal exponent $\gamma$, which for $d\rightarrow\infty$ is $\gamma=2.34\ldots$~\cite{kurchan2013exact}. Remarkably, over the accessible dynamical range, both models roughly follow that scaling, with agreement improving as $d$ increases. (Deviations in $\gamma$ have the opposite sign as what has been reported for (standard) hard spheres~\cite{charbonneau2022dimensional}, but given the non-universality of the exponent and the absence of perturbative prediction for it little can be concluded from this difference.) Second, consider the peak height. MFT  suggests~\cite{franz2011field,biroli2022local}
\begin{equation}
\hat \alpha_2(\hat t_\mathrm{peak}) \sim |\hat\varphi_\mathrm{d} - \hat\varphi|^{-1}.
\end{equation}
Upon approaching $\hat\varphi_\mathrm{d}$, both models appear to follow a master curve given by that scaling.  These pseudo-critical scalings are therefore consistent with an (avoided) critical transition at $\hat\varphi_\mathrm{d}$ in finite $d$. Unfortunately, the relevant regime is too short to detect systematic deviation of  criticality below the predicted (perturbative) upper critical dimension, $d_\mathrm{u}=8$~\cite{biroli2007critical,franz2011field,franz2012field}, as has been reported for related exponents~\cite{berthier2020finite}. 

Deviations in peak height from the expected pseudo-critical scaling are more revealing. For both systems, the effect gradually decrease as $d$ increases. Remarkably, while for the MK model deviations steadily drift up, for the RLG deviations plateau more sharply. The difference likely reflects the complete absence of cooperativity in the latter system, but whether this effect has to do with perturbative or instantonic processes is not immediately obvious. To see more clearly, we leverage the broad $d$ range accessible for the RLG to determine that the plateau then grows roughly linearly with $d$ (Fig.~\ref{fig:a2peakscaling} inset). The peak divergence of $\alpha_2$ is therefore suppressed as $1/d$, hence suggesting that the plateau height is perturbatively controlled, albeit at a higher order in the expansion.

\section{Susceptibility time evolution and long-time scaling}
\label{sec:chi4}

\begin{figure*}
\includegraphics[width=1\textwidth]{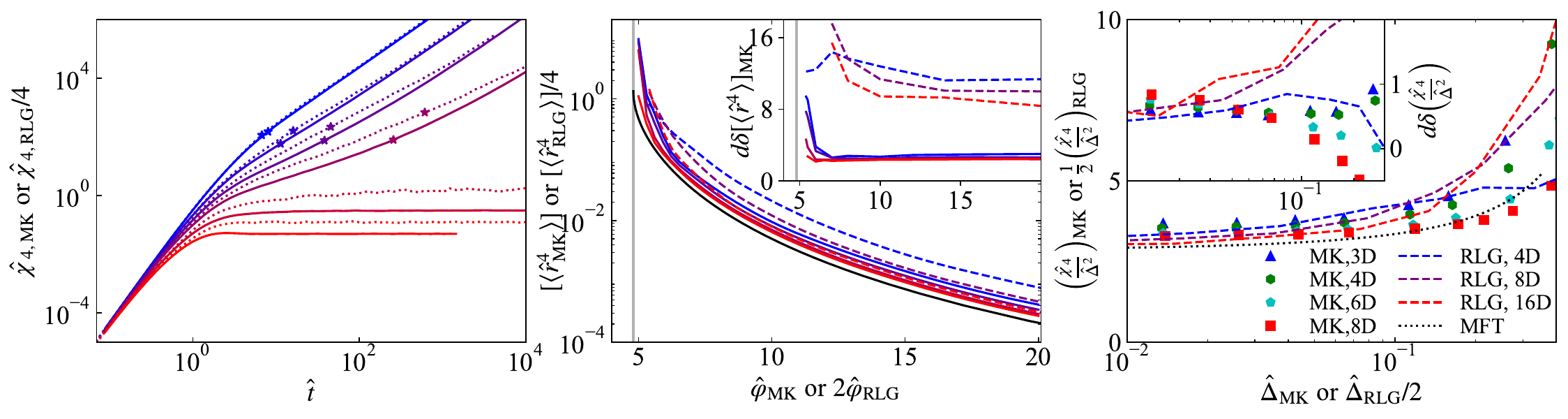}
\caption{(a) Time evolution of $\hat\chi_4$ for $\hat\varphi_\mathrm{MK}=2 \hat\varphi_\mathrm{RLG}=1,2,3,4,6,8$ (from blue to red lines) for the MK model in $d=8$ (solid lines) and for the RLG in $d=24$ (dashed lines). The peak of $\hat\alpha_2$ from Fig.~\ref{fig:a2peakscaling} (asterisk) identifies the intermediate time regime in the diffusive regime.
(b) In the caging regime, the mean quartic displacement for the MK model in $d=3,4,6,8$ (blue to red solid lines) and the RLG in $d=4,8,16$ (blue to red dashed lines) approaches the MFT prediction (black) as $d$ increases. (inset) Both systems show perturbative $1/d$ corrections, $\delta [\hat r^4(\hat{t})] = |([\hat r^4_\mathrm{MFT}(\hat{t})] - [\hat r^4(d,\hat{t})])/[\hat r^4_\mathrm{MFT}(\hat{t})]$, but those are markedly larger in the RLG than in the MK model.
(c) In the caging regime, long-time $\hat\chi_4/\hat\Delta^2$ in the MK model (markers) and the RLG (from Ref.~\cite{biroli2022local}, dashed lines) converge to the  $d\rightarrow\infty$ prediction (black dotted line). (inset) The relative deviation from the MFT prediction in the form of Eq.~\eqref{eq:chi4corr} is consistent with a perturbative correction deep in the caging regime.}
\label{fig:chi4compare}
\end{figure*}

Consistent with the $1/d$ corrections to the MSD differing for the RLG and the MK model (see Fig.~\ref{fig:brmsdshorttime})
$\hat\alpha_{2}$ quantitatively differs between the two systems,
but this difference cannot be explained by the simple mapping relations given by Eqs.~\eqref{eq:mapping2} and \eqref{eq:a2chimap} (Fig.~\ref{fig:mkalpha2time}). As expressed by Eq.~\eqref{eq:a2chimap}, $\hat\alpha_2$ depends solely on the MSD and $\hat\chi_4$ for a given $d$. Given that the former is fairly close to the DMFT prediction for both the RLG and the MK model in all $d$, the main difference in $\alpha_2$ presented in Sec.~\ref{sec:alpha2} must therefore come from the latter of the two quantities.

The time evolution of $\hat\chi_4$ for the RLG and the MK model is compared in Fig.~\ref{fig:chi4compare}(a) for the highest available $d$ for each system. As expected from Eq.~\eqref{eq:mapping2}, the trivial (factor of 4) mapping collapses the two sets of curves at short times. In the caging regime, the additional multiplicative factor from Eq.~\eqref{eq:mapping2} separates the two sets of results at long times. In the diffusive regime, Eq.~\eqref{eq:a2chimap} gives that $\hat\alpha_2(\hat t \rightarrow \infty) = 0$, and hence $\hat\chi_4 = 2 \hat\Delta^2$ at long times. Both curves then again converge.  At intermediate times, where $\alpha_2$ peaks, however, results for the two systems markedly deviate. 

To tease out the origin of this difference, we expand the MSD and the mean quartic displacement perturbatively in $1/d$ (see Figs.~\ref{fig:msdnewbr}, \ref{fig:mkplateau}, \ref{fig:chi4compare}(b))
\begin{equation} \label{eq:mfrexpansion}
\begin{aligned}
\ [\langle \hat r^2(\hat t) \rangle] &= \hat\Delta_\mathrm{MF}(\hat t) + A_1(\hat t) / d + A_2 (\hat t) / d^2 + \ldots \\
[\langle \hat r^4(\hat t) \rangle] &= [\langle \hat r^4(\hat t) \rangle]_\mathrm{MF} + B_1(\hat t) / d + B_2 (\hat t) / d^2 + \ldots \, ,
\end{aligned}
\end{equation}
Because all $d$ dependencies have been extracted for the MF values in Eq.~\eqref{eq:mfrexpansion}, and hence $[\langle \hat r^4 \rangle]_\mathrm{MF} = \hat\Delta_\mathrm{MF}^2$, we then have
\begin{equation} \label{eq:chi4expand}
\left( \hat \chi_4 / \hat\Delta^2 \right) (\hat t) = \frac{B_1 - 2 A_1 \hat\Delta_\mathrm{MF} }{\hat\Delta_\mathrm{MF}^2} + O(1/d),
\end{equation}
where for notational clarity the time dependence is omitted on the right-hand side. Through Eq.~\eqref{eq:a2chimap}, $\hat\alpha_2(t)$ can then be described \emph{to leading order} by the \emph{leading perturbative} correction of the MSD and the mean quartic displacement (see Figs.~\ref{fig:msdnewbr},~\ref{fig:mkplateau}, and~\ref{fig:chi4compare}(b) insets), which are not universal.

Because quantitative predictions exist for $\chi_4/\Delta^2$ in the limit $d\rightarrow\infty$, both the dimensional convergence of $\hat\chi_4$ can also be evaluated. 
As shown in Fig.~\ref{fig:chi4compare}(c), a solid quantitative agreement is obtained for small cages (i.e. large $\hat\varphi$) in both models, with corrections overall scaling as $1/d$ at large $\hat\varphi$,
\begin{equation} \label{eq:chi4corr}
\delta \left( \frac{\hat \chi_4 }{ \hat\Delta^2} \right) = \left(\left( \frac{\hat \chi_4 }{ \hat\Delta^2} \right) - \left( \frac{\hat \chi_4 }{ \hat\Delta^2} \right)_\mathrm{MFT}\right)/ \left( \frac{\hat \chi_4 }{ \hat\Delta^2} \right)_\mathrm{MFT} \sim 1/d.
\end{equation}
Upon approaching $\hat\varphi_\mathrm{d}$, however, the growth of $\hat\chi_4/\hat\Delta^2$ deviates from the MFT prediction. For the RLG, it was shown in Ref.~\cite{biroli2022local} that this deviation is due to the presence of weak cages, as captured in the large tail in the distribution of cages around $\hat\varphi_\mathrm{d}$. For the MK model cages are stronger and more narrowly distributed than for the RLG,  as discussed above. The results therefore more closely trail the $d\rightarrow\infty$ prediction. Results for $d=8$, however, are somewhat confounding. We suspect that deviations around $\hat\varphi_\mathrm{d}$ might be due to finite-$d$ corrections associated with the difference between mean and modal quantities, but identifying the optimal finite-$d$ estimator for $\chi_4$ in this regime is left for future studies.

\section{Conclusion}
\label{sec:conclusion}

Using state-of-the-art numerical simulations, we have studied the fluctuations of two minimally structured glass formers in finite $d$, thus identifying the non-Gaussian 
parameter $\alpha_2$ as a perturbative $1/d$ correction to the exact $d\rightarrow\infty$ DMFT. The observable is therefore intimately related to mean-field--like caging and cage escapes, and not solely to (non-perturbative) hopping or jumping physics, as is sometimes suggested~\cite{berthier2011theoretical}. 
Given the physical centrality and dimensional robustness of the numerical results for $\alpha_2$, a description of small fluctuations around the DMFT in the diffusive regime would be a significant enrichment of the theory of simple glasses. The simulation results presented in this work can serve both as targets for a future such calculation and, in the meantime, as reference for the study of single-particle fluctuations in (standard) hard sphere glass formers. Such fluctuation-based comparison would offer a much more stringent test of the MFT for model glass formers than has thus far been possible.
 
More generally, the overall success of this work motivates further pursuing the program of extracting the out-of-equilibrium DMFT description from finite-$d$ simulations of minimally structured model glass formers. Insight into the mean-field--like features that survive in $d=3$ system will then be more readily obtained than has thus far been possible by solving the associated equations.

\acknowledgments
We thank  G.~Folena, F.~Ricci-Tersenghi, A.~Manacorda and F.~Zamponi for various stimulating discussions, and the latter two also for sharing updated DMFT results for Fig.~\ref{fig:brmsdshorttime}. Computations were carried out on the Duke Compute Cluster. This work was supported by a grant from the Simons Foundation (Grant No. 454937). Data relevant to this work have been archived and can be accessed at Duke digital repository at https://doi.org/10.7924/xxxxxxxx.


\appendix

\section{Simulation boxes} \label{appd:box}

Simulations are generally run in boxes under checker-board $D_d$ periodic boundary conditions, as described in Refs.~\cite{charbonneau2020percolation,charbonneau2022dimensional}. This setup enables a system size reduction of a factor of $2^{(d-2)/2}$ relative to standard (hyper)cubic boundary conditions $Z_{d}$ for similar finite-size corrections from the thermodynamic limit. For the MK model, in which the number of pairs of coordinate shifts, $\OO(N^2)$, sets the memory complexity, simulations up to $d=8$ are then numerically accessible. The chosen system size aims to balance computational cost and finite-size corrections, such that the MSD is not affected significantly ($\lesssim 1\%$ systematic error). We here find that $N=3000$ suffices for all cases except for $\hat\varphi > \hat\varphi_\mathrm{d}$ in $d=8$. For instance, $N=10,000$ is needed for $\hat\varphi = 20$, thus limiting the computationally accessible range of densities.

The single-particle nature of the RLG model allows for the dimensional range to stretch to $d=12$. That range can be further expanded by using periodic boundary conditions with the $d=24$ Leech lattice symmetry, $\Lambda_{24}$~\cite{conway1986soft,cohn2017sphere}. This exceptionally dense arrangement allows for a reduction in system size by up to a factor $2^{24}$ and $2^{13}$ relative to a system under with $Z_{24}$ (hypercubic) and $D_{24}$ symmetry, respectively~\cite{charbonneau2022dimensional}. Despite this remarkable efficiency gain, we argued in Ref.~\cite{charbonneau2022dimensional} that the decoding cost of $\Lambda_{24}$ was nevertheless too onerous to make it computationally practical. However, a more efficient decoder devised by Vardy~\emph{et al.}~\cite{vardy1993maximum,van2016cryptographic} has since come to our attention. That decoder requires only 3,595 operations (versus 55,968 for that of Ref.~\cite{conway1986soft}) to identify the minimal image of a particle, and thus makes it computationally accessible.

The resulting implementation provides roughly $1.2\times10^5$ queries-per-second (QPS) performance for the minimal image computation (tested on Intel Xeon E5-2680 v3 CPU single core). It then takes four hours to run a $d=24$ system with $N=10^6$ up to $\hat t=10^5$. For reference, running the MK model in $d=8$ with $N=3000$ up to $\hat t=10^5$ takes three days on that same architecture.

\section{Microscopic Dynamics}  \label{appd:microdynamics}

\subsection{Newtonian Dynamics}

A standard event-driven implementation of Newtonian dynamics is used for both the RLG and the MK model~\cite{skoge2006packing,charbonneau2014hopping}. 
For the RLG, however, a subtle correction also needs to be taken into account. Earlier numerical simulations used a unit tracer velocity throughout~\cite{biroli2020unifying,charbonneau2020percolation,biroli2022local}. This microcanonical ensemble setup does not recover the (Gaussian) Maxwell-Boltzmann distribution of velocities of an equilibrium multi-particle system. As a result, the self-van Hove function is then non-Gaussian at short times~\cite{biroli2022local}.
Although such microscopic details of the dynamics are expected to be irrelevant at long times upon approaching $\hat\varphi_\mathrm{d}$ and beyond, the resulting short-time deviations are sufficiently significant  to partly obfuscate the emergence of interesting features at intermediate times. In order to sidestep this difficulty, we here assign the tracer an initial velocity taken from a Gaussian distribution, so as to recover the Maxwell-Boltzmann distribution of velocities. The tracer velocity (preserving its direction) is also reassigned with probability $p = 1/d$ at each collision. Because the tracer collision rate (number of collisions per unit time) scales as $\hat Z_\mathrm{tracer} \sim d \hat\varphi$ (derived in the following), this choice results in a $d$-independent scaling of the decorrelation time from the initial velocity rate that matches that of multi-particle models, such as standard and MK hard spheres. In addition, the long-time dynamics of the system remains unaffected.

\subsubsection{Collision rate scaling}

The collision rate of the RLG tracer can be derived using a fairly straightforward kinetic theory analysis. Suppose a tracer colliding at the origin and moving freely to $r$. The probability that the tracer has a free run up to $r$ is given by the probability that obstacles are absent within a cylinder with top area of a half unit sphere and length $r$. This shell has volume $V_\mathrm{shell} = V_{d-1} r$. For Poisson distributed obstacles, the cumulative probability that the first collision happens at range $(0, r)$, $Q_\mathrm{collision}(r)$, then satisfies the relation
\begin{equation}
Q_\mathrm{collision} (r) = 1 - \exp(-\rho V_{d-1} r).
\end{equation}
The average free distance available to the tracer is therefore
\begin{equation}
\bar r = \int_0^\infty \dv{Q_\mathrm{collision}(r)}{r} r \dd r = \frac{1}{\rho V_{d-1}},
\end{equation}
and its collision rate, with the time unit scaled as $\hat{t} = \sqrt{d} t$ is
\begin{equation} \label{eq:Zpredict}
\hat Z_\mathrm{tracer} = 1/(\sqrt{d} \bar r) = (d\hat\varphi /\sqrt{\pi d}) \frac{\Gamma(1+\frac{d}{2})}{\Gamma(\frac{1+d}{2})}.
\end{equation}
Recall that $\rho V_d = d\hat\varphi$. We then obtain the large $d$ limit scaling of $\hat Z_\mathrm{tracer}$ as
\begin{equation} \label{eq:Zscaling}
\hat Z_\mathrm{tracer} \sim \frac{d \hat\varphi}{\sqrt{2 \pi}}.
\end{equation}

\begin{figure}
\centering
\includegraphics[width=0.5\textwidth]{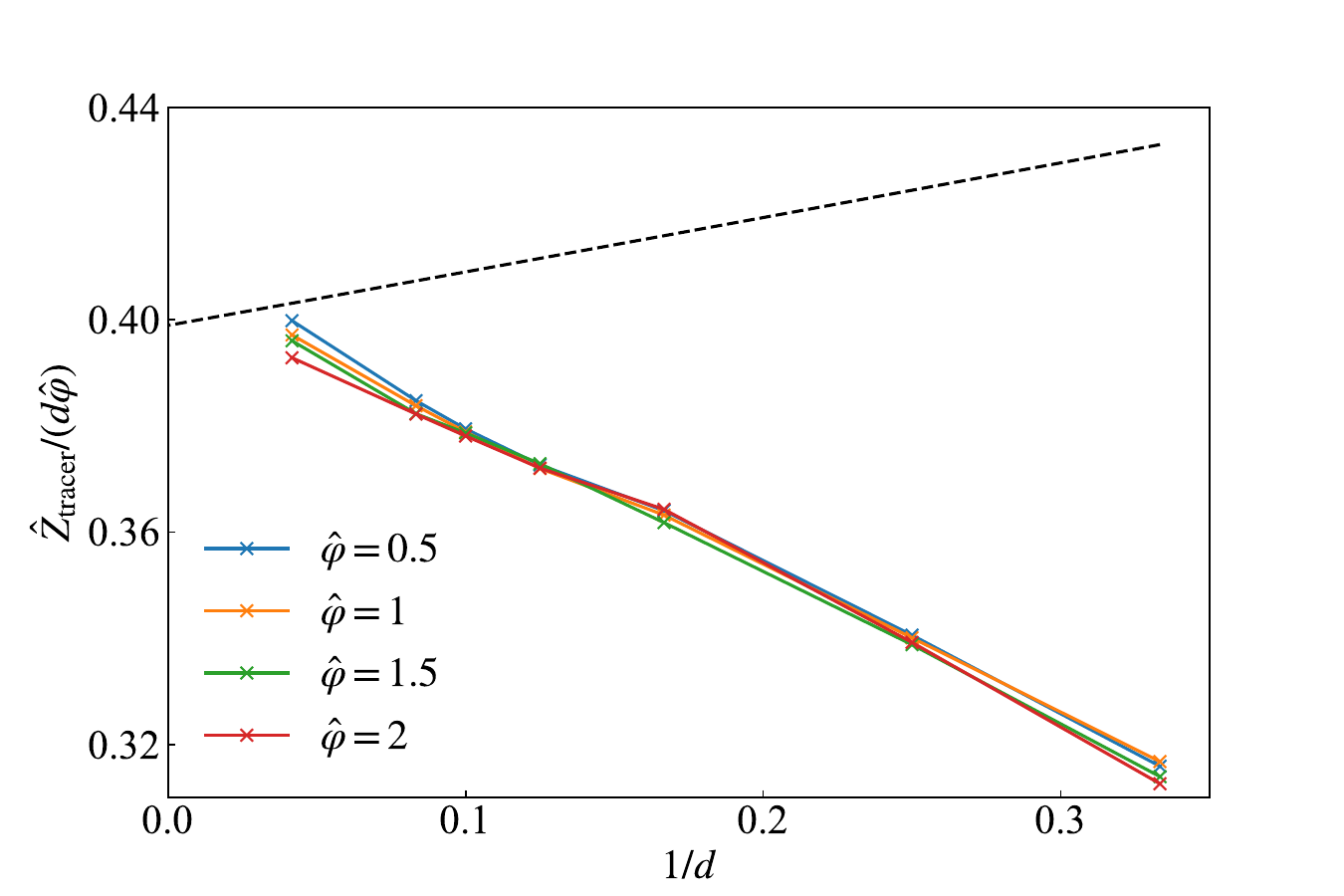}
\caption{Scaling of the tracer collision rate $\hat Z_\mathrm{tracer}$. Crosses are simulation results for various obstacle densities. Black dashed line denotes the relation of Eq.~\eqref{eq:Zpredict}.}
\label{fig:collrate}
\end{figure}

Figure~\ref{fig:collrate} shows the numerical results for $\hat Z_\mathrm{tracer}$. The $d,\hat\varphi$ scaling and the large $d$ limit are all consistent with Eq.~\eqref{eq:Zscaling}. The $1/d$ correction, however, has the opposite sign as that predicted by Eq.~\eqref{eq:Zpredict}. Correlations between collisions, which are neglected in the above treatment and grow with $\hat\varphi$, are likely the origin of this discrepancy. Calculation of the associated $1/d$ correction, however, is left for future consideration.

\subsection{Brownian Dynamics}

For the RLG, Brownian dynamics is also considered, using de Michele's event-driven scheme~\cite{foffi2005scaling,scala2007event}. In this scheme, at the end of every time interval $\Delta \hat t_n = 2^n \Delta \hat t_0 \le 1/2$ particle velocity is reset to a multi-variate Gaussian random variable, $\bm{v}= \bm{g}/\sqrt{\Delta t_n}$. As minimal time interval, we choose $\Delta \hat t_0 = \hat t_\mathrm{min} /10$, where $\hat t_\mathrm{min}$ is the smallest sampling time interval of correlations; after sampling $\hat t_\mathrm{min}$ for $2^{10}$ times (at $\hat t=2^{10} \hat t_\mathrm{min}$), we set $n=1$; after sampling $2 \hat t_\mathrm{min}$ for $2^{10}$ times (at $\hat t=2^{11} \hat t_\mathrm{min}$), we set $n=2$; and so on until $\Delta \hat t_n \le 1/2$. This strategy recovers the Brownian statistics at short times, while avoiding the excessive computational cost of repeatedly re-generating random velocities and recalculating subsequent collisions over longer timescales, where this effect plays no notable role.

\section{Non-universal mapping of susceptibility} \label{appd:mapping}

From Eq.~\eqref{eq:mapping}, one expects 
\begin{equation} \label{eq:chimapping2}
\hat\chi_\mathrm{4,RLG} \leftrightarrow 4 \hat\chi_\mathrm{4,MK},
\end{equation}
but this expression only holds at short times and in the long-time diffusive limit ($\hat\varphi < \hat\varphi_\mathrm{d}$), in which cases the tracer displacement is Gaussian by construction. In the long-time caging limit, although Eq.~\eqref{eq:mapping} still holds, Eq.~\eqref{eq:chimapping2} does not. In that regime, $\hat\chi_4$ is non-zero as a result of $1/d$ corrections from the mean-field result to $[ \langle \hat{r}^4(t) \rangle ]$ and $\hat\Delta$ (as discussed in Sec.~\ref{sec:chi4}). 

A mapping is also possible in the long-time caging limit. Recall that in  Ref.~\onlinecite[Eqs.~(S11-13)]{charbonneau2014hopping} the mean-field cage size of the MK model was computed from a Gaussian random variable of variance $A_\mathrm{MK}$, which measures the cage size along one direction with $\Delta = 2 d A_{\mathrm{MK}}$.
Reference~\cite{biroli2020unifying} further noted that the large variance term has the form $\hat\Delta = (\hat \Delta_\mathrm{tracer} + \hat \Delta_\mathrm{obstacle}) / 2$, where for the MK model $\Delta_\mathrm{tracer}$ and $\Delta_\mathrm{obstacle}$ are equivalent, whereas in the RLG the obstacles are pinned, i.e., $\Delta_\mathrm{obstacle} = 0$.
As a result, $\Delta_\mathrm{RLG}$ and $\Delta_\mathrm{MK}$ involve $d$ and $2d$ independent random variables, respectively. Therefore,
\begin{equation} \label{eq:deltamapping}
\begin{aligned}
\Delta_\mathrm{RLG} &= d A_\mathrm{RLG}, \\
\Delta_\mathrm{MK} &=d (A_\mathrm{MK,tracer} + A_\mathrm{MK,obstacle}) = 2 d A_\mathrm{MK}.
\end{aligned}
\end{equation}
Because $\Delta_\mathrm{RLG} \leftrightarrow 2 \Delta_\mathrm{MK}$, we have $A_\mathrm{RLG} \leftrightarrow 4 A_\mathrm{MK}$, and the contribution of the variance of the cage size due to single random variable is $\mathrm{Var}(A_\mathrm{RLG}) \leftrightarrow 16 \mathrm{Var}(A_\mathrm{MK})$.
Equation~\eqref{eq:deltamapping} also gives that the expected variance is
\begin{equation}
\begin{aligned}
\mathrm{Var}(\Delta_\mathrm{RLG}) &= d \mathrm{Var}(A_\mathrm{RLG}), \\
\mathrm{Var}(\Delta_\mathrm{MK}) &= 2 d \mathrm{Var}(A_\mathrm{MK}). \\
\end{aligned}
\end{equation}
We then have $\mathrm{Var}(\Delta_\mathrm{RLG}) \leftrightarrow 8 \mathrm{Var}(\Delta_\mathrm{MK})$, i.e.,
\begin{equation}
\hat\chi_\mathrm{4,RLG} \leftrightarrow 8 \hat\chi_\mathrm{4,MK}.
\end{equation}
More physically, the additional factor of 2 relative to Eq.~\eqref{eq:chimapping2} comes from the different number of independent terms in the variance. In the RLG only the tracer is oscillating within the cage, whereas in the MK model both the tracer and the surrounding obstacles are random variables that contribute to caging.

\section{Short-time expansion of DMFT equation for Brownian dynamics}
\label{appd:brshorttime}

For Brownian hard spheres, the DMFT equation (Ref.~\cite[Eq,~3]{manacorda2020numerical}) reads
\begin{equation} \label{eq:dmftbrhs}
\pdv{\hat \Delta}{\hat t} = 1- \int_0^{\hat t} \dd u \mathcal{M}_\mathrm{HS}(\hat t-u) \pdv{\hat \Delta}{u}.
\end{equation}
where $\mathcal{M}_\mathrm{HS}(t)$ is the memory function. We here seek a short-time analytical expansion form of $\hat \Delta(\hat t)$ from this equation.

Replacing $\mathcal{M}_\mathrm{HS}$ with the first-order approximation of $\hat\varphi$ (Ref.~\cite[Eq.~24]{manacorda2020numerical}),
\begin{equation} \label{eq:memoryexpand}
\begin{aligned}
\mathcal{M}_\mathrm{HS}(\hat t) &= \hat\varphi \mathcal{M}_\mathrm{HS}^{(1)}(\hat t) + \OO(\hat\varphi^2) \\
&= \frac{\hat\varphi}{2} \left [ \frac{\exp(-\hat t/4)}{\sqrt{\pi \hat t}} - \frac{\erfc(\sqrt{\hat t}/2)}{2} \right ] + \OO(\hat\varphi^2), \\
&= \frac{\hat\varphi}{2} \left [ \frac{1}{\sqrt{\pi \hat t}} - \frac{1}{2} + \OO({\hat t}^{\frac{1}{2}}) \right ] + \OO(\hat\varphi^2),
\end{aligned}
\end{equation}
and assuming
\begin{equation}
\pdv{\hat \Delta}{\hat t} = 1 - A(\hat\varphi) {\hat t}^{\frac{1}{2}} + B(\hat\varphi) \hat t + \OO(t^2),
\end{equation}
we can solve for the integral equation~\eqref{eq:dmftbrhs} and evaluate the coefficients
\begin{equation}
\begin{cases}
A &= \hat\varphi / \sqrt{\pi} \\
B &= \frac{\hat\varphi}{4}(1 + \hat\varphi).
\end{cases}
\end{equation}

\begin{figure}
\centering
\includegraphics[width=0.5\textwidth]{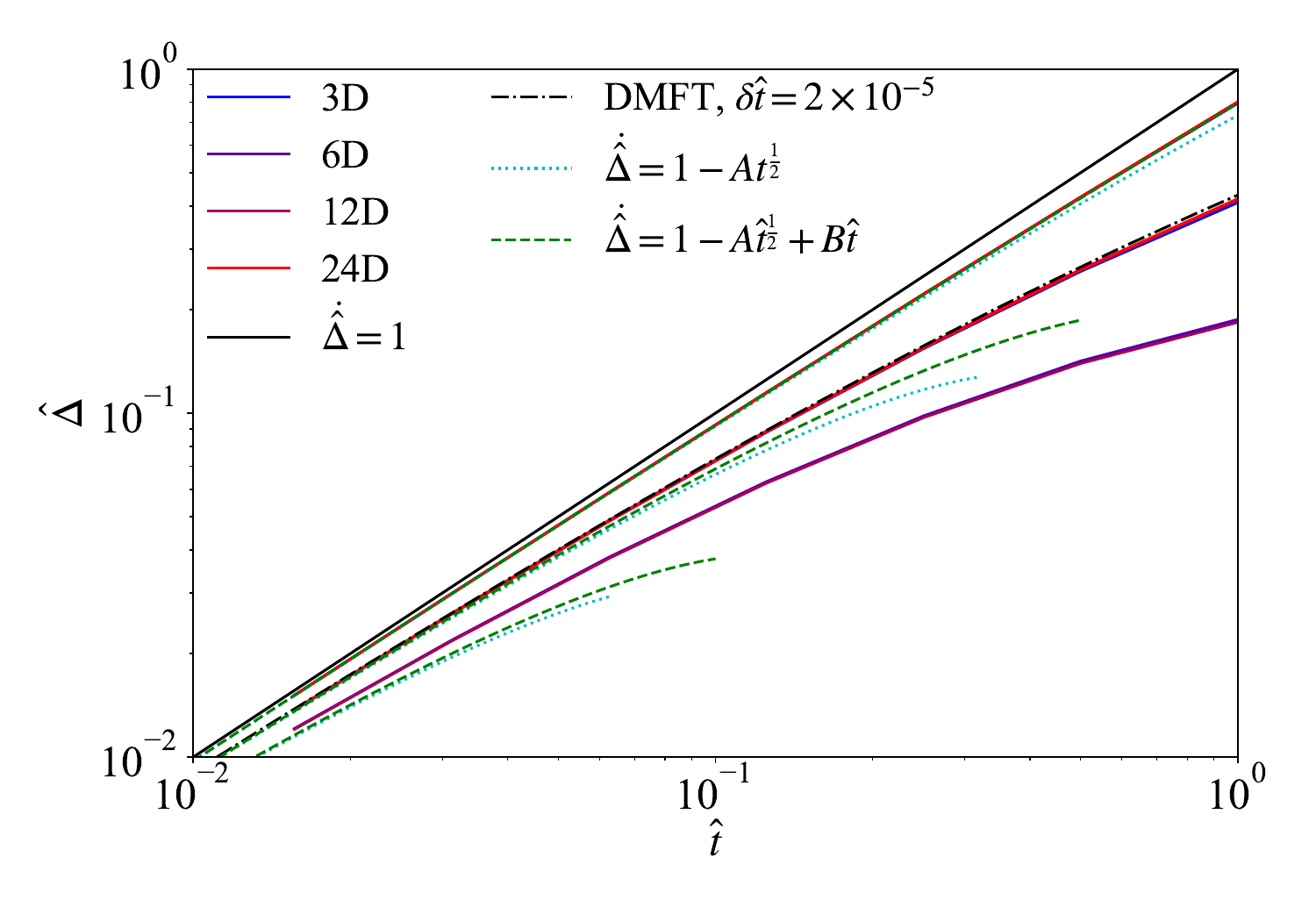}
\caption{Brownian dynamics of RLG at short time for $\hat\varphi=0.5,2,4$ (from top to bottom), compared with the DMFT result of smaller integration step ($\delta t=2 \times 10^{-5}$) as well as the expansion forms of Eq.~(\ref{eq:brmsdshorttime}, \ref{eq:brmsdshorttime2}).}
\label{fig:brmsdshorttime}
\end{figure}

Note that we only keep linear terms of $\hat\varphi$ in $A, B$ because Eq.~\eqref{eq:memoryexpand} is expanded to that order.
We then have
\begin{equation}  \label{eq:brmsdshorttime}
\hat\Delta_\mathrm{HS} (t) = \hat t - \frac{2}{3} \frac{\hat\varphi_\mathrm{HS}}{\sqrt{\pi}} {\hat t}^{\frac{3}{2}} + \frac{\hat\varphi_\mathrm{HS}}{8} {\hat t}^2 + \OO(\hat t^3).
\end{equation}
For the RLG, the mapping gives
\begin{equation} \label{eq:brmsdshorttime2}
\hat\Delta_\mathrm{RLG} (t) = \hat t - \frac{2\sqrt{2}}{3} \frac{\hat\varphi_\mathrm{RLG}}{\sqrt{\pi}} {\hat t}^{\frac{3}{2}} + \frac{\hat\varphi_\mathrm{RLG}}{8} {\hat t}^2 + \OO(\hat t^3).
\end{equation}

The short-time simulation results are fully consistent with Eq.~\eqref{eq:brmsdshorttime2} (Fig.~\ref{fig:brmsdshorttime}). The expansion results also highlight that noticeable corrections---up to the sub-dominant correction term ($B \hat t$)---becomes increasingly pronounced at higher densities. The short-time discrepancy in Fig.~\ref{fig:collrate}(b) is therefore clearly due to numerical accuracy issues with solving Eq.~\eqref{eq:dmftbrhs}. In particular, setting a smaller numerical integration step $\delta \hat t=2 \times 10^{-5}$ (instead of $\delta \hat t=10^{-2}$ in Ref.~\cite{manacorda2020numerical}) results a much closer agreement with simulations.

\end{document}